\documentclass[12pt]{article}
\usepackage{bbm,latexsym,epsfig}
\newcommand{\mnu}{\mathcal{M}_\nu}
\newcommand{\deltasol}{\Delta m^2_\odot}
\newcommand{\deltaatm}{\Delta m^2_\mathrm{atm}}
\newcommand{\berr}{\!\begin{array}{l} \scriptstyle +}
\newcommand{\tr}{\\[-3mm] \scriptstyle -}
\newcommand{\eerr}{\end{array}}

\newcommand{\yi}{h'_{33}}

\newcommand{\yiii}{g'_{23}}
\newcommand{\yiv}{g'_{12}}  
\newcommand{\yv}{f'_{22}}
\newcommand{\yvi}{f'_{33}}
\newcommand{\yvii}{f'_{11}}
\newcommand{\yviii}{f'_{13}}

\begin{document}

\title{\normalsize \hfill UWThPh-2006-15 \\[1cm]
\LARGE
Fermion masses and mixings \\
in a renormalizable $SO(10) \times \mathbbm{Z}_2$ GUT \\[8mm]}

\author{
Walter Grimus\thanks{E-mail: walter.grimus@univie.ac.at} \,
\normalsize and \large
\setcounter{footnote}{3}
Helmut K\"uhb\"ock\thanks{E-mail: helmut.kuehboeck@gmx.at}
 \\
\small Institut f\"ur Theoretische Physik, Universit\"at Wien \\
\small Boltzmanngasse 5, A--1090 Wien, Austria
\\*[4.6mm]}

\date{20 September 2006}

\maketitle

\begin{abstract} 
We investigate a scenario 
in a supersymmetric $SO(10)$ Grand Unified Theory
in which the fermion mass matrices 
are generated by renormalizable Yukawa couplings 
of the $\mathbf{10} \oplus \mathbf{120} \oplus \overline{\mathbf{126}}$ 
representation of scalars.
We reduce the number of parameters by 
assuming spontaneous CP violation and a 
$\mathbbm{Z}_2$ family symmetry, leading to nine real Yukawa coupling
constants for three families.
Since in the ``minimal SUSY $SO(10)$ GUT'' an intermediate seesaw scale is
ruled out and our scenario lives in the natural extension of this theory  
by the $\mathbf{120}$, we identify the vacuum
expectation value (VEV) $w_R$ of 
$(\mathbf{10}, \mathbf{1}, \mathbf{3}) \in \overline{\mathbf{126}}$
with the GUT scale of $2 \times 10^{16}$ GeV. 
In order to obtain sufficiently large neutrino masses, the coupling matrix of
the scalar $\overline{\mathbf{126}}$ is necessarily small and we neglect 
type II seesaw contributions to the light-neutrino mass matrix.
We perform a numerical analysis of this 21-parameter scenario 
and find an excellent fit to experimentally known fermion masses and mixings. 
We discuss the properties of our numerical solution, including a 
consistency check for the VEVs of the Higgs-doublet components in the 
$SO(10)$ scalar multiplets.
\end{abstract}

\newpage

\paragraph{Introduction:}
The group $SO(10)$ is a favourite candidate for constructing grand
unified theories (GUTs)~\cite{fritzsch}. The special interest in such
theories also stems from the fact that they allow for 
type~I~\cite{seesaw} and type~II~\cite{typeII} seesaw mechanisms 
(see also~\cite{seesaw-general})
for the light neutrino masses.
Confining oneself to renormalizable $SO(10)$ GUTs,
the scalar representations coupling to the chiral fermion
fields, which are all assembled for each family 
in the 16-dimensional irreducible representation (irrep), 
are determined by the relation~\cite{sakita,slansky}
\begin{equation}
\mathbf{16} \otimes \mathbf{16}
= \left( \mathbf{10} \oplus \mathbf{126} \right)_\mathrm{S}
\oplus \mathbf{120}_\mathrm{AS},
\label{tensor}
\end{equation}
where the subscripts ``S'' and ``AS'' denote, respectively,
the symmetric and antisymmetric parts of the tensor product.
The so-called ``minimal SUSY $SO(10)$ GUT'' (MSGUT)~\cite{aulakh83}
makes use of one $\mathbf{10}$
and one $\overline{\mathbf{126}}$ scalar irrep 
for the Yukawa couplings, 
to account for all fermion masses and mixings~\cite{babu92}. 
The MSGUT contains, in addition, one
$\mathbf{210}$ and one $\mathbf{126}$ scalar irrep~\cite{aulakh83}.
This model 
has built-in the gauge-coupling unification
of the minimal SUSY extension of the Standard Model (MSSM).
Detailed studies of this minimal theory have been
performed~\cite{fukuyama99,fukuyama01,okada,bajc02,%
goh,bertolini,malinsky,macesanu}; 
in~\cite{fukuyama99,bertolini,malinsky} 
small effects of the 120-plet were considered in addition. 
It turned out that the MSGUT works surprisingly well in the fermion
sector, provided one neglects constraints on the overall scale of the light
neutrino masses. This, however, proved to be crucial, since the
natural order of the neutrino masses in GUTs is too low, namely 
$v^2/M_\mathrm{GUT} \sim 1.5 \times 10^{-3}$ eV, with 
$v\sim 174$ GeV and the GUT scale 
$M_\mathrm{GUT} \sim 2 \times 10^{16}$ GeV. 
Thorough studies of the 
heavy scalar states~\cite{AG02,fukuyama04,melfo,AG04,fukuyama,aulakh05} 
have been used to show that this MSGUT
is too constrained~\cite{au05,bajc05} 
and does not allow to enhance the neutrino mass scale to a
realistic one~\cite{garg,schwetz}, compatible with the results of the
neutrino oscillation experiments 
(for a review see, e.g.,~\cite{nu-review}). 
One aspect of this problem is that a seesaw scale significantly lower than the
GUT scale spoils the gauge coupling unification of the MSSM.

An obvious attempt to loosen the corset of the minimal theory is to add the
120-plet of scalars. 
A study in that direction has been done
in~\cite{aulakh06}. Earlier works considering a prominent 120-plet
contribution to the fermion mass matrices are found 
in~\cite{oshimo,yang,dutta1,dutta2}.
We note that $\mathbf{10} \oplus \mathbf{120}$
alone does not give a good fit in the charged fermion
sector~\cite{LKG}. Thus the $\overline{\mathbf{126}}$ scalar irrep is
not only needed in the neutrino sector but also for the charged
fermion mass matrices.
In that case, the mass matrices of 
the charged fermions 
and the neutrino Dirac-mass matrix
are given, respectively, by
\begin{eqnarray}
M_d    & = & k_d\, H + \kappa_d\,    G +   v_d\, F, 
\label{md} \\
M_u    & = & k_u\, H + \kappa_u\,    G +   v_u\, F, 
\label{mu} \\
M_\ell & = & k_d\, H + \kappa_\ell\, G - 3 v_d\, F, 
\label{ml} \\
M_D    & = & k_u\, H + \kappa_D\,    G - 3 v_u\, F.
\label{mD}
\end{eqnarray}
The Yukawa coupling matrices $H$, $G$, $F$ belong to the 
scalar irreps 
$\mathbf{10}$, $\mathbf{120}$, $\overline{\mathbf{126}}$,
respectively. The coefficients $k_d$, $\kappa_d$, $\kappa_\ell$, $v_d$ 
denote the vacuum expectation values (VEVs) of the Higgs doublet
components in the respective $SO(10)$ scalar 
irreps which contribute to the MSSM Higgs doublet $H_d$, the
rest of the coefficients refers to $H_u$. 
The light neutrino mass matrix is obtained as
\begin{equation}\label{mnu}
\mnu = M_L - M_D M_R^{-1} M_D^T
\quad \mbox{with} \quad
M_L = w_L\, F, \quad
M_R = w_R\, F, 
\end{equation}
with scalar triplet VEVs $w_L$ and $w_R$.
The mass Lagrangian of the ``light'' fermions reads 
\begin{equation}
\mathcal{L}_M  = 
- \bar d_L M_d\, d_R - \bar u_L M_u\, u_R - \bar \ell_L M_\ell\, \ell_R -
\frac{1}{2} \bar\nu_L \mnu \left( \nu_L \right)^c + \mbox{H.c.},
\end{equation}
with $\left( \nu_L \right)^c$ being the charge-conjugate of $\nu_L$.

\paragraph{A renormalizable $SO(10)$ scenario:}
The goal of this letter is a numerical study of the 
system of 3-generation mass matrices~(\ref{md}) to (\ref{mnu}),
taking into account the neutrino-mass suppression factor 
$v^2/M_\mathrm{GUT}$.
This system does not easily lend itself to such an investigation
because it contains many parameters, thus we use some 
arguments to reduce their number.
The scenario we want to investigate is defined by the following assumptions:
\begin{enumerate}
\renewcommand{\labelenumi}{\roman{enumi})}
\item
The Yukawa coupling matrices $H$, $G$, $F$ are real.
\item
We impose a $\mathbbm{Z}_2$ symmetry, which sets some of the Yukawa
couplings to zero and which is spontaneously broken by 
the VEVs of the $\mathbf{120}$, in particular, by 
$\kappa_d$, $\kappa_u$, $\kappa_\ell$, $\kappa_D$ being non-zero.
\item
We assume $w_R = M_\mathrm{GUT}$, with 
$M_\mathrm{GUT} = 2 \times 10^{16}$ GeV. 
\item
We set $w_L = 0$, i.e., we have pure type~I seesaw mechanism.
\end{enumerate}
Let us now comment on these items.
Item i) can be motivated by spontaneous CP violation. The
$\mathbbm{Z}_2$ of item ii) is given by
\begin{equation}\label{2}
\psi_2 \to -\psi_2, \quad 
\phi_\mathbf{120} \to -\phi_\mathbf{120},
\end{equation}
where the $\psi_j$ ($j=1,2,3$) denote the fermionic 16-plets and 
$\phi_\mathbf{120}$ is the scalar 120-plet. All other multiplets, not
mentioned in Eq.~(\ref{2}), transform trivially.
With the $\mathbbm{Z}_2$ symmetry of Eq.~(\ref{2}), 
the coupling matrices have the form
\begin{equation}\label{Y}
H = \left(
\begin{array}{ccc}
h_{11} & 0 & 0 \\ 0 & h_{22} & 0 \\ 0 & 0 & h_{33}
\end{array} \right), \quad
G = \left(
\begin{array}{ccc}
0 & g_{12} & 0 \\ -g_{12} & 0 & g_{23} \\ 0 & -g_{23} & 0
\end{array} \right), \quad
F = \left(
\begin{array}{ccc}
f_{11} & 0 & f_{13} \\ 0 & f_{22} & 0 \\ f_{13} & 0 & f_{33}
\end{array} \right).
\end{equation}
We have used the freedom of basis choice in the 1--3 sector 
to set $h_{13} = 0$.
Of course, this $\mathbbm{Z}_2$ symmetry of Eq.~(\ref{2}) is an ad-hoc
symmetry, but it enhances the importance of the $\mathbf{120}$ because
its Yukawa coupling matrix $G$ is now responsible for mixing of the
second family with the other two.\footnote{In Eq.~(\ref{2}), 
for the definition of the $\mathbbm{Z}_2$ symmetry, 
all choices $\psi_j \to -\psi_j$ are equivalent. With choosing
$\psi_2$, we anticipate the result of the fit of our scenario to the
masses and mixings at the GUT scale. That fit gives a strong
hierarchy of the elements of $H$, which---with Eq.~(\ref{2})---can be
formulated in the usual way as
$\left| h_{11} \right| \ll \left| h_{22} \right| \ll 
\left| h_{33} \right|$. Furthermore, with the convention of~(\ref{2})
it is possible to have all diagonalizing matrices of the charged fermion
masses in the vicinity of the unit matrix.}
Item iii) is motivated by the fact that 
the MSGUT does not allow to fix the problem of too small neutrino
masses by taking $w_R$ significantly lower 
than the GUT scale~\cite{au05,bajc05,garg,schwetz,aulakh06}. Thus our
scenario has built in that the natural neutrino mass scale in the
MSGUT is too low. Consequently, the neutrino mass scale has to be
enhanced by the smallness of the coupling matrix $F$~\cite{aulakh06}.
Item iv) is a trivial consequence of the previous one: for small $F$,
type II seesaw contribution to $\mnu$ is negligible.

Now we tackle the problem of parameter counting. Without loss of generality,
we assume that $k_d$, $k_u$ and $w_R$ are real and positive. Then we define
\begin{equation}\label{M'}
H' = k_d H, \quad G' = \left| \kappa_d \right| G, \quad
F' = \left| v_d \right| F.
\end{equation}
The primed matrices have the dimension of mass. The phases of the VEVs of the
120 and 126-plets cannot be removed. Thus we write the mass matrices as 
\begin{eqnarray}
M_d    & = & H' + e^{i\xi_d} G' + e^{i\zeta_d} F', 
\label{Md} \\
M_u    & = & r_H H' + r_u\, e^{i\xi_u} G' + r_F e^{i\zeta_u} F', 
\label{Mu} \\
M_\ell & = & H' + r_\ell\, e^{i\xi_\ell} G' - 3\, e^{i\zeta_d} F', 
\label{Ml} \\
M_D    & = & r_H H' + r_D\, e^{i\xi_D} G' - 3\, r_F e^{i\zeta_u} F', 
\label{MD} \\
\mnu   & = & r_R\, M_D {F'}^{-1} M_D^T.
\label{Mnu} 
\end{eqnarray}
The ratios $r_H$, etc., are real by definition since we have extracted
the phases from the VEVs.
Now the counting is easily done. Since we have nine real Yukawa
couplings, see Eq.~(\ref{Y}),
there are nine real parameters in $H'$, $G'$, $F'$. Furthermore, there
are six phases and six (real) ratios of VEVs, 
altogether 21 real parameters.
On the other hand, we have 18 observables we want to fit: nine charged-fermion
masses, three mixing angles and one CP phase in the CKM matrix, two
neutrino mass-squared differences $\deltaatm$ and $\deltasol$, and three
lepton mixing angles.

Suppose, we have obtained a good fit for the 18 observables. 
Then we still have to
check if the fit allows for reasonable VEVs and Yukawa coupling constants.
A detailed discussion of this issue is found in Appendix A. Here it is
sufficient to note that $w_R = M_\mathrm{GUT}$ and the determination of $r_R$
and $r_F$ by the fit fix  
$\left| v_d \right|$ and $\left| v_u \right|$ via
$\left| v_d \right| = r_R M_\mathrm{GUT}$ and 
$\left| v_u \right| =  r_F \left| v_d \right|$.
Therefore, as a first test we check
\begin{equation}\label{1sttest}
\left| v_d \right|^2 + \left| v_u \right|^2 =
\left| v_d \right|^2 \left( 1 + r_F^2 \right) < v^2 
\quad \mbox{with} \quad v = 174\: \mathrm{GeV}
\end{equation}
for every fit. Clearly, this inequality holds at the electroweak
scale, and we assume that approximately it is valid at the GUT scale too.

\begin{table}[t]
\begin{center}
\renewcommand{\arraystretch}{1.2}
\begin{tabular}{cc}
\begin{tabular}[t]{|c|c|} \hline
\multicolumn{2}{|c|}{Quarks} \\ \hline\hline
$m_d$ &  
$1.5036 \berr 0.4235 \tr 0.2304 \eerr$ \\ \hline
$m_s$ & $29.9454 \berr 4.3001 \tr 4.5444 \eerr$ \\ \hline
$m_b$ & $1063.6 \berr 141.4 \tr 086.5 \eerr$ \\ \hline
$m_u$ & $0.7238 \berr 0.1365 \tr 0.1467 \eerr$ \\ \hline
$m_c$ & $210.3273 \berr 19.0036 \tr 21.2264 \eerr$ \\ \hline
$m_t$ & $82433.3 \berr 30267.6 \tr 14768.6 \eerr$ \\ \hline
$s_{12}$ & $0.2243 \pm 0.0016$ \\ \hline
$s_{23}$ & $0.0351 \pm 0.0013$ \\ \hline
$s_{13}$ & $0.0032 \pm 0.0005$ \\ \hline
$\delta_{CKM}$ & $60^\circ \pm 14^\circ$ \\ \hline
\end{tabular}
&
\begin{tabular}[t]{|c|c|} \hline
\multicolumn{2}{|c|}{Leptons} \\ \hline\hline
$m_e$       &  
$0.3585 \berr 0.0003 \tr 0.0003 \eerr$ \\ \hline
$m_\mu$     & 
$75.6715 \berr 0.0578 \tr 0.0501 \eerr$ \\ \hline
$m_\tau$    & 
$1292.2 \berr 0.0013 \tr 0.0012 \eerr$ \\ \hline
$\deltasol$ & $(7.9 \pm 0.3) \times 10^{-5}$ \\ \hline
$\deltaatm$ & $\Big(2.2 \berr 0.37 \tr 0.27 \eerr 
\Big) \times 10^{-3}$ \\ \hline
$s_{12}^2$  & $0.31 \pm 0.025$ \\ \hline
$s_{23}^2$  & $0.50 \pm 0.065$ \\ \hline
$s_{13}^2$  & $< 0.0155$ \\ \hline
\end{tabular}
\end{tabular}
\end{center}
\caption{Input data at the GUT scale for $M_\mathrm{GUT} = 2 \times
  10^{16}$ GeV and $\tan \beta = 10$. The charged-fermion masses are
  taken from~\cite{das}, the remaining input from Table~I
  in~\cite{schwetz}. Charged-fermion masses are in units of MeV,
  neutrino mass-squared differences in eV$^2$. We have used the
  abbreviations $s_{12} \equiv \sin \theta_{12}$, etc. 
  The angles in the left table refer to the CKM matrix, in the right
  table to the PMNS matrix.\label{input}}
\end{table}
\paragraph{A numerical solution:}
To find a numerical solution, we employ the downhill simplex
method~\cite{downhill} for
minimizing a $\chi^2$-function of the parameters---for an
explanation of the method see~\cite{schwetz,LKG}.
Actually, the $\chi^2$-function can be minimized analytically with respect to
the parameter $r_R$ of Eq.~(\ref{Mnu}), which results in a
$\chi^2$-function depending the remaining 20 parameters, and we apply our
numerical method to that function. To build in the
inequality~(\ref{1sttest}) in our search for the minimum, we add
a suitable penalty function to our $\chi^2$.
Our scenario is fitted against the values of the 18 observables at the
GUT scale; for an MSSM parameter $\tan \beta = 10$, these values are
displayed in Table~\ref{input}.

Choosing the normal ordering $m_1 < m_2 < m_3$ of the neutrino masses 
($\deltasol = m_2^2 - m_1^2$, $\deltaatm = m_3^2 - m_1^2$),
we have found a fit with a $\chi^2 = 0.0087$, 
which is a perfect fit for all practical purposes.
This fit is so good that it does not make sense to show the
pulls.\footnote{The largest pull is $5 \times 10^{-2}$ for $m_s$.}
The matrices $H'$, $G'$ and $F'$ for our fit are given by
\begin{eqnarray}
H' & = & \left( \begin{array}{ccc}
0.716986 & 0 & 0 \\ 0 & -40.6278 & 0 \\ 0 & 0 & 1114.41
\end{array} \right), \nonumber \\
G' & = & \left( \begin{array}{ccc}
0 & 7.56737 & 0 \\ -7.56737 & 0 & 36.8224 \\ 0 & -36.8224 & 0 
\end{array} \right),
\label{solution} \\
F' & = & \left( \begin{array}{ccc}
-0.0966851 & 0 & 4.25282 \\ 0 & 12.3136 & 0 \\ 4.25282 & 0 & -61.6491 
\end{array} \right), \nonumber
\end{eqnarray}
where all numerical values are in units of MeV; the values of the
ratios of VEVs and the phases are shown in Table~\ref{ratios+phases}. 
\begin{table}
\begin{center}
\renewcommand{\arraystretch}{1.2}
\begin{tabular}{|c|c||c|c|} \hline
$r_H$    & 91.0759   & -          & - \\ \hline
 -       & -       & $\zeta_d$  & $19.66974^\circ$ \\ \hline
$r_F$    & 297.758   & $\zeta_u$  & $-2.96594^\circ$ \\ \hline
 -       & -       & $\xi_d$    & $189.12385^\circ$ \\ \hline
$r_u$    & 7.14572   & $\xi_u$    & $226.65689^\circ$ \\ \hline
$r_\ell$ & 1.33897   & $\xi_\ell$ & $6.24258^\circ$ \\ \hline
$r_D$    & 3008.88 & $\xi_D$    & 
$179.85271^\circ$ \\ \hline
$r_R$    & $2.90553 \times 10^{-17}$ & - & - \\ \hline
\end{tabular}
\end{center}
\caption{The values of the phases and ratios appearing in the mass
  matrices~(\ref{Md})--(\ref{Mnu}) in the case of our 
  fit. Hyphens in the left two columns indicate that the
  ratio corresponding to the phase has been absorbed in one of the primed
  matrices, whereas hyphens in the right two columns signify that there is no
  physical phase associated with that ratio. \label{ratios+phases}} 
\end{table}
The neutrino mass spectrum turns out to be hierarchical with
$m_1 = 1.57 \times 10^{-3}\: \mbox{eV} \ll 
m_2 = 9.03 \times 10^{-3}\: \mbox{eV} \ll m_3 = 46.96 \times
10^{-3}\: \mbox{eV}$,
and the PMNS phase\footnote{We use the same phase convention as for
  the CKM matrix in~\cite{RPP}.} is $12^\circ$.
We want to stress, however, that our fit solution
is perhaps not unique, because with the numerical method used here we  
could miss other minima of $\chi^2$. 

\begin{figure}[t]
\begin{center}
\epsfig{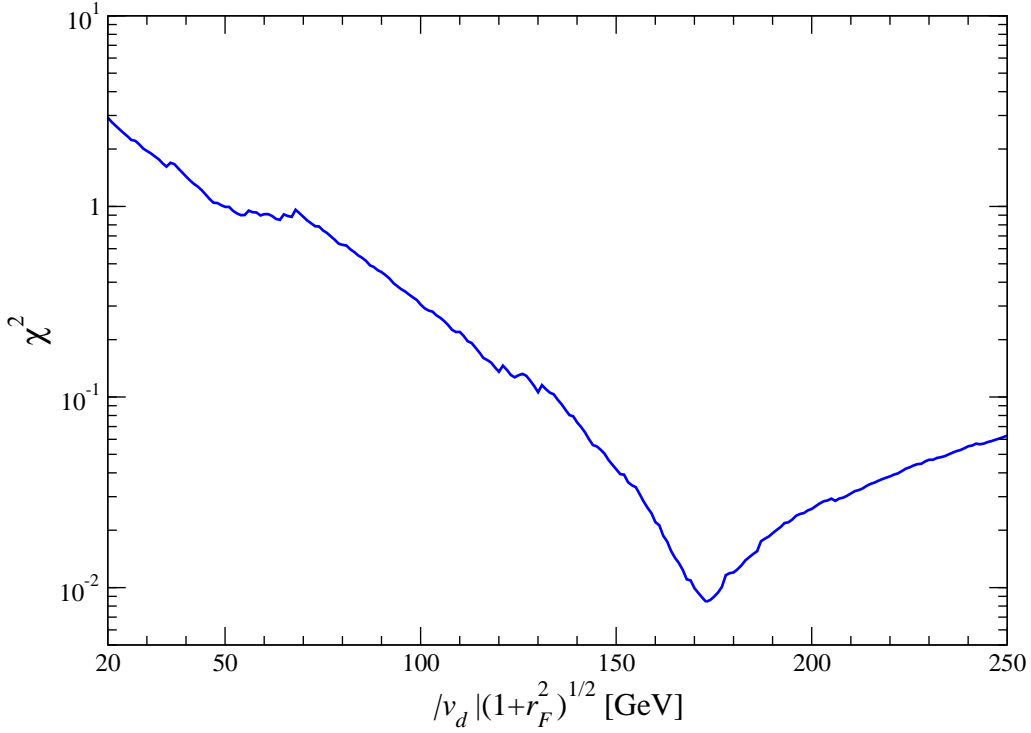}
\end{center}
\caption{The $\chi^2$ as a function of 
  $\left| v_d \right| \sqrt{1 + r_F^2}$. \label{vev}}
\end{figure}
For our fit, it turns out that 
$y \equiv \left| v_d \right| \sqrt{1 + r_F^2} = 173.0$ GeV. This looks
dangerously close to the upper bound of Eq.~(\ref{1sttest}). To check
if this danger is serious, 
we have plotted in Fig.~\ref{vev} the minimal $\chi^2$ as a function of
$y$. In order to pin $y$
down to a given value $\bar y$ we have extended the $\chi^2$ function to
$(\chi^2)_y = \chi^2 + \left\{ (y - \bar y)/(0.01 \bar y) \right\}^2$, 
minimized $(\chi^2)_y$ and plotted $\chi^2$ at this minimum versus
$\bar y$---for previous uses of this method 
see, for instance,~\cite{schwetz}.
We read off from Fig.~\ref{vev} that $\chi^2$ is minimal at $y = 173$
GeV, however, this minimum is rather flat; note that $\chi^2$ is
plotted on a logarithmic scale. Thus we still obtain excellent fits if
we go to lower values of $y$. In Appendix~A, a consistency condition 
is worked out which the $SO(10)$ GUT has to fulfill in order to
reproduce the VEV ratios of Table~\ref{ratios+phases}. 
There we also show that for our fit all Yukawa couplings stay in the
perturbative regime. 

In order to find out if our scenario makes a
prediction for the PMNS phase $\delta$, we treat it in the same
way as $y$ in the previous paragraph, i.e., we consider 
$(\chi^2)_\delta = \chi^2 + 
\left\{ (\delta - \bar \delta)/(0.01 \bar \delta) \right\}^2$.
Departing from $\bar \delta = 12^\circ$ for our numerical solution
given by Eq.~(\ref{solution}) and Table~\ref{ratios+phases}, and going
stepwise down close to $\bar\delta = 0^\circ$ and up to 
$\bar\delta = 360^\circ$, the quality of the fits remains excellent,
with $(\chi^2)_\delta$ always below 0.3.
Thus, in our scenario all values of the PMNS phase are possible.

One may ask the question 
how large enough neutrino masses and an atmospheric
mixing angle which is close to maximal are accomplished with the numerical
values given by Eq.~(\ref{solution}) and Table~\ref{ratios+phases}.
We concentrate on achieving $\sqrt{\deltaatm} \sim 0.05$ eV. With the value of
$r_R$ we find that
$r_R \times 10^9\:\mbox{MeV} \simeq 0.029 \:\mbox{eV}$. Thus we take
into account all
contributions to $\mnu/r_R$ which are of order $10^9$ MeV. Rewriting
Eq.~(\ref{Mnu}) in the form
\begin{equation}
\mnu = r_R \left\{ \left( r_H H' + r_D\, e^{\xi_D} G' \right) 
{F'}^{-1} \left( r_H H' + r_D\, e^{i\xi_D} G' \right)^T -
6\, r_H r_F\, e^{i\zeta_u} H' + 9\, r_F^2\, e^{2i\zeta_u} F'
\right\},
\end{equation}
a numerical analysis shows that all elements of the second and third term on
the right-hand side are smaller than $1.8 \times 10^8$ MeV and
$0.5 \times 10^8$ MeV, respectively. 
Thus the dominant matrix elements in $\mnu$
stem from the first term and are given by
\begin{equation}\label{dom}
\mnu^{(\mathrm{dom})} = r_R\, r_D \left(
\begin{array}{ccc}
0 & 0 & 0 \\ 
0 &
r_D \left( 2 \yiii \yiv \yviii + \yvi \left( \yiv \right)^2 \right)/d &
-r_H \yi \yiv \yviii/d \\
0 & 
-r_H \yi \yiv \yviii/d &
r_D \left( \yiii \right)^2/\yv
\end{array} \right).
\end{equation}
Here, $d = \yvii \yvi - \left( \yviii \right)^2$ is the determinant of the 
corresponding $2 \times 2$ submatrix of $F'$ and we have used the
approximation 
$\xi_D = 180^\circ$. 
Apart from the common factor $r_R$, 
in this matrix there are four products of three matrix 
elements: one matrix element is always from ${F'}^{-1}$, 
the other two are either
both from $r_D G'$ or one from $r_D G'$ and one from $r_H H'$.
Looking at Eq.~(\ref{solution}) and Table~\ref{ratios+phases}, we find that 
products of the largest elements, 
for instance $\left( \yiii \right)^2 \yvi$,
never occur, these would be too large. 
Plugging in the numerical values of the parameters, we find that 
all non-zero terms in $\mnu^{(\mathrm{dom})}$ 
are similar in magnitude:
\begin{equation}
r_R \times 10^9\, \mbox{MeV} \,
\left( \begin{array}{cc}
(-1.77 + 2.64) & -0.81 \\ -0.81 & 1.00
\end{array} \right).
\end{equation}
Due to the minus sign in the first term, we end up with 
$-1.77 + 2.64 = 0.87$, close to $1.00$ and thus leading to nearly
maximal mixing. In this crude approximation, which is only relevant for
the largest mass and the atmospheric mixing angle, we obtain 
$m_3 \simeq 0.05$ eV and $\theta_{23} \simeq 43^\circ$.\footnote{Note that
  there is also a small contribution from $M_\ell$.} Apart from the
smallness of $F'$, it is the large factor $r_D$ in $M_D$ which gives the
correct magnitude of the neutrino masses. Maximal atmospheric neutrino mixing
rather looks like a numerical contrivance in our scenario.

Finally, a word concerning the low-energy values of the quark masses
is at order. Ref.~\cite{das} takes the quark mass values at $m_Z$ from
Ref.~\cite{fusaoka} as input for the renormalization group evolution
up to $M_\mathrm{GUT}$, whereas Ref.~\cite{fusaoka} uses the input 
\begin{equation}\label{1GeV}
m_u(1\,\mbox{GeV}) = 4.88 \pm 0.57, \quad
m_d(1\,\mbox{GeV}) = 9.81 \pm 0.65, \quad
m_s(1\,\mbox{GeV}) = 195.4 \pm 12.5
\end{equation}
and
\begin{equation}
m_c(m_c) = 1.302 
\begin{array}{c} 
\scriptstyle +0.037 \\[-1.5mm] \scriptstyle -0.038 
\end{array}\!, \quad
m_b(m_b) = 4.34 
\begin{array}{c} 
\scriptstyle +0.07 \\[-1.5mm] \scriptstyle -0.08 
\end{array}\!, \quad
m_t(m_t) = 171 \pm 12,
\end{equation}
see Tables~I and II in~\cite{fusaoka}. 
The light quark masses are given in MeV, the heavy ones in GeV.
Comparing these values with those given in the Review of Particle
Properties of 2006 (RPP)~\cite{RPP}, 
we find that the heavy quark masses are in reasonable agreement. 
However, in the last years the values of the light quark
masses have become significantly lower~\cite{RPP}:
\begin{equation}\label{2GeV}
m_u(2\,\mbox{GeV}) = 1.5 \div 3.0, \quad
m_d(2\,\mbox{GeV}) = 3 \div 7, \quad
m_s(2\,\mbox{GeV}) = 95 \pm 25,
\end{equation}
Note that one has to take into account the scaling factor
$m_i(1\,\mbox{GeV})/m_i(2\,\mbox{GeV}) = 1.35$ ($i = u,\,d,\,s$) 
to compare Eq.~(\ref{1GeV}) with Eq.~(\ref{2GeV})~\cite{RPP}.
In order to assess the influence of lowering the light quark masses, we
have performed a second fit, using the values of Eq.~(\ref{2GeV}) as
input, scaled to $M_\mathrm{GUT}$ with the factor 0.200 for $m_u$ and
0.207 for $m_d$ and $m_s$ (see~\cite{das,fusaoka}), but
leaving the previous values for the heavy quark masses. We found an
excellent fit with $\chi^2 = 0.052$, which means that our scenario is
able to reproduce the lower values of the light quark masses as well.
The second fit has some qualitative differences in comparison with the
first one, which reinforces the suspicion that, for given input values
of the 18 observables, the fit solution in our scenario is not unique.

\paragraph{Summary:}
In this paper we have investigated fermion masses and mixings in the
$SO(10)$ MSGUT, augmented by a 120-plet of scalars.
The main purpose was to show that in this setting it 
is possible to reconcile the type I seesaw mechanism (see Eq.~(\ref{mnu})) 
with a triplet VEV $w_R$ equal to the GUT scale of $2 \times 10^{16}$ GeV,
provided the theory admits that the MSSM Higgs doublet $H_d$ is composed
mainly of the corresponding doublet components in the 
$\mathbf{126}$ and $\mathbf{210}$ scalar irreps---see Eq.~(\ref{norm2}); 
those are the irreps which have no Yukawa couplings. 
This reconciliation was feasible within the scenario defined in points
i)--iv), in which we have used
symmetries to significantly reduce the number of degrees of freedom in
the Yukawa couplings---see Eq.~(\ref{Y}).
Within this scenario we were able to find an
excellent fit for all fermion masses and mixings; in this fit we have 
a hierarchical neutrino mass spectrum.\footnote{We have also tried
  fits for the inverted ordering. In that case, the best fit we found
  has $\chi^2 = 1.8$ and $m_3 \simeq 7 \times 10^{-6}$ eV.}

Thus we have obtained the following results for the minimal
renormalizable $SO(10)$ GUT, with Yukawa couplings according to the
relation~(\ref{tensor}):
\begin{itemize}
\item
It is possible to reproduce the 
correct neutrino mass scale.
\item
Nevertheless, gauge coupling unification is not spoiled.
\item
The concrete $SO(10)$ scenario with type I seesaw mechanism, 
we treated in this paper, has 
21 parameters, just as the MSGUT with type I+II seesaw mechanism and 
general complex Yukawa couplings.
\end{itemize}

It remains to be studied if our scenario allows a sufficient
suppression of proton decay. In~\cite{dutta2} it was shown that the
scalar 120-plet plays a crucial role for that purpose; a certain 
texture of the Yukawa coupling matrices---similar to our numerical
solution~(\ref{solution})---enables that suppression even 
for large $\tan \beta$. 

\vspace{5mm}

\noindent
\textit{Acknowledgments:}
W.G. thanks C.S.\ Aulakh for illuminating discussions and L.\ Lavoura
for reading the manuscript.

\appendix
\setcounter{equation}{0}
\renewcommand{\theequation}{A\arabic{equation}}

\section{The MSSM Higgs doublets and the mass matrices}

The MSSM contains two Higgs doublets, $H_d$ and $H_u$, with hypercharges $+1$
and $-1$, respectively. Their corresponding VEVs are denoted by 
$v \cos \beta$ and $v \sin \beta$ ($v = 174$ GeV), respectively. 
Neglecting effects of the electroweak scale, these
doublets are, by assumption, the only scalar zero modes extant at the GUT
scale; this requires a minimal finetuning condition~\cite{AG02,melfo}. 
The scalar irreps
$\mathbf{10}$, $\overline{\mathbf{126}}$, $\mathbf{126}$, $\mathbf{210}$
contain each one doublet with the quantum numbers of $H_d$, whereas the
$\mathbf{120}$ contains two such doublets. The $H_d$ is composed of these
doublets~\cite{melfo} with the corresponding amplitudes~\cite{garg} 
$\bar\alpha_j$ ($j = 1,\ldots,6$). 
The analogous coefficients for $H_u$ are denoted by
$\alpha_j$. The normalization conditions are
\begin{equation}\label{normalization}
\sum_{j=1}^6 \left| \alpha_j \right|^2 = 
\sum_{j=1}^6 \left| \bar\alpha_j \right|^2 = 1.
\end{equation}
The Dirac mass matrices, taking into account that the $\mathbf{126}$ and
$\mathbf{210}$ have no Yukawa couplings, are given by 
\begin{eqnarray}
M_a & = &
v \cos \beta \left( 
c^a_1 \bar\alpha_1 Y_{10} + c^a_2 \bar\alpha_2 Y_{\overline{126}} +
      \left( c^a_5 \bar\alpha_5 + c^a_6 \bar\alpha_6 \right) Y_{120} \right)
\quad (a = d, \ell), \label{Ma} \\
M_b & = &
v \sin \beta \left( 
c^b_1 \alpha_1 Y_{10} + c^b_2 \alpha_2 Y_{\overline{126}} +
      \left( c^b_5 \alpha_5 + c^b_6 \alpha_6 \right) Y_{120} \right)
\quad (b = u, D), \label{Mb}
\end{eqnarray}
with Yukawa coupling matrices $Y_{10}$, $Y_{\overline{126}}$, $Y_{120}$ and
Clebsch-Gordan coefficients $c^{a,b}_j$ deriving from the $SO(10)$-invariant
Yukawa couplings~\cite{AG04,au05}. The absolute values 
of the Clebsch-Gordan coefficients have no physical meaning and some
of their phases are convention-dependent.
With our conventions, the required information reads
\begin{equation}\label{c}
\begin{array}{l}
c^d_1 = c^u_1 = c^\ell_1 = c^D_1, \\
c^d_2 = -c^u_2 = -\frac{1}{3}\,c^\ell_2 = \frac{1}{3}\,c^D_2, \\
c^d_5 = -c^u_5 = c^\ell_5 = -c^D_5, \\
c^d_6 = c^u_6 = -\frac{1}{3}\,c^\ell_6 = -\frac{1}{3}\,c^D_6, \\
c^d_5/c^d_6 = \sqrt{3}.
\end{array}
\end{equation}
Equations~(\ref{Ma}) and (\ref{Mb}) together with this equation lead
to the mass matrices~(\ref{md})--(\ref{mD}). Furthermore, 
comparing Eqs.~(\ref{Ma}) and (\ref{Mb}) with Eq.~(\ref{M'}), we find
\begin{equation}
H' = v \cos \beta\, c^d_1 \left| \bar\alpha_1 \right| Y_{10}, 
\quad
F' = v \cos \beta\, c^d_2 \left| \bar\alpha_2 \right| Y_{\overline{126}}, 
\quad
G' = v \cos \beta 
\left| c^d_5 \bar\alpha_5 + c^d_6 \bar\alpha_6 \right| Y_{120}.
\end{equation}
Comparison with Eqs.~(\ref{Md})--(\ref{MD}) and using Eq.~(\ref{c})
delivers the coefficients
\begin{eqnarray}
r_H = \tan \beta \left| \frac{\alpha_1}{\bar\alpha_1} \right|, &
\displaystyle r_F = \tan \beta \left| 
\frac{\alpha_2}{\bar\alpha_2} \right|, & \label{r1} \\
r_u = \tan \beta \left| 
\frac{\alpha_6 - \sqrt{3}\, \alpha_5}{\bar\alpha_6 - \sqrt{3}\,
  \bar\alpha_5} \right|, & \displaystyle
r_\ell = \left| 1 - \frac{2\,\bar\alpha_6}{\bar\alpha_6 - \sqrt{3}\,
  \bar\alpha_5} \right|, & 
r_D = \tan \beta \left| 
\frac{3\,\alpha_6 + \sqrt{3}\, \alpha_5}{\bar\alpha_6 - \sqrt{3}\,
  \bar\alpha_5} \right|.
\label{r2}
\end{eqnarray}

Now we want to check the consistency of our numerical solution given by
Eq.~(\ref{solution}) and Table~\ref{ratios+phases}. 
From $r_D \gg r_u$, it follows that 
\begin{equation}
\sqrt{3}\,\alpha_5 \simeq \alpha_6 \simeq \frac{r_D}{4\,\tan \beta}
\left| \bar\alpha_6 - \sqrt{3}\, \bar\alpha_5 \right|.
\end{equation}
Furthermore, using $r_\ell \sim 1$, we find the order-of-magnitude relations
\begin{equation}
\bar\alpha_5 \sim \bar\alpha_6 \sim \tan \beta/r_D.
\end{equation}
Then the first of the normalization conditions~(\ref{normalization}) reads
approximately 
\begin{equation}
\sum_j \left| \alpha_j \right|^2 \simeq 
\frac{1}{\tan^2 \beta} \left( r_H^2 \left| \bar\alpha_1 \right|^2 + 
r_F^2 \left| \bar\alpha_2 \right|^2 + 
\frac{1}{12}\, r_D^2 
\left| \bar\alpha_6 - \sqrt{3}\, \bar\alpha_5 \right|^2 \right) + 
\left| \alpha_3 \right|^2 + \left| \alpha_4 \right|^2 \simeq 1.
\end{equation}
This means that $\left| \bar\alpha_j \right|^2 \ll 1$ for $j = 1,2,5,6$.
Therefore, the second normalization condition is given by 
\begin{equation}\label{norm2}
\sum_j \left| \bar\alpha_j \right|^2 \simeq
\left| \bar\alpha_3 \right|^2 + \left| \bar\alpha_4 \right|^2 \simeq 1,
\end{equation}
and the brunt of the normalization has to be supplied by the components of
$H_d$ in the $\mathbf{126}$ and $\mathbf{210}$, 
which do not couple to the fermions. This is a consistency condition for 
the scenario presented in this paper. 

To translate the condition~(\ref{1sttest}) into the formalism
presented here, we note that 
$\left| v_d \right|^2 \ll \left| v_u \right|^2$ and
$\sin \beta \simeq 1$ for $\tan\beta = 10$. Therefore, Eq.~(\ref{1sttest})
effectively checks if the necessary condition $\left| \alpha_2 \right| < 1$ is
fulfilled. 

Finally, it remains to see if our numerical solution respects the
perturbative regime of the Yukawa sector. It suffices to consider the largest
elements of the Yukawa couplings
\begin{equation}
Y_d    = \frac{1}{v \cos\beta}\,M_d,    \quad
Y_u    = \frac{1}{v \sin\beta}\,M_u,    \quad
Y_\ell = \frac{1}{v \cos\beta}\,M_\ell, \quad
Y_D    = \frac{1}{v \sin\beta}\,M_D,
\end{equation}
which reside in $Y_u$ and $Y_D$. The largest entry in $Y_u$ is the 33-element
with the main contribution from $r_H \yi/(v \sin \beta) \simeq 0.59$. 
The 23-element with $r_D \yiii/(v \sin \beta) \simeq 0.64$ dominates in $Y_D$.
These numbers demonstrate that for our numerical solution the Yukawa couplings
remain in the perturbative regime.

\end{document}